# 3D Printed Metallic Dual-Polarized Vivaldi Arrays on Square and Triangular Lattices

Carl Pfeiffer, Jeffrey Massman, and Thomas Steffen

*Abstract*— We report the first Vivaldi arrays monolithically fabricated exclusively using commercial, low-cost, 3D metal printing (direct metal laser sintering). Furthermore, we developed one of the first dual-polarized Vivaldi arrays on a triangular lattice, and compare it to a square lattice array. The triangular lattice is attractive because it has a 15.5% larger cell size compared to the square lattice and can be more naturally truncated into a wide range of aperture shapes such as a rectangle, hexagon, or triangle. Both arrays operate at 3-20 GHz and scan angles out to 60º from normal. The fabrication process is significantly simplified compared to previously published Vivaldi arrays since the antenna is ready for use directly after the standard printing process is complete. This rapid manufacturing is further expedited by printing the "Sub-Miniature Push-on, Micro" (SMPM) connectors directly onto the radiating elements, which simplifies assembly and reduces cost compared to utilizing discrete RF connectors. The arrays have a modular design that allow for combining multiple sub-arrays together for arbitrarily increasing the aperture size. Simulations and measurement show that our arrays have similar performance as previously published Vivaldi arrays, but with simpler fabrication.

## I. INTRODUCTION

Active electronic scanning arrays (AESAs) with ultra-wide bandwidths are appealing for space constrained platforms because they can be used for a multitude of missions such as radar, communication, and direction finding all within a single aperture. A myriad of ultra-wideband (UWB) antenna geometries have been investigated which offer tradeoffs between bandwidth, polarization purity, fabrication complexity, and efficiency. Some examples include tightly coupled dipoles and slots [1, 2], Planar Ultrawideband Modular antenna (PUMA) [3, 4], Balanced Antipodal Vivaldi Array (BAVA) [5, 6, 7], Frequency-scaled Ultra-wide Spectrum Element (FUSE) [8], and Sliced Notch Array [9]. Vivaldi antennas (also known as notch antennas) are particularly attractive since they are simple to design and can offer a good impedance match over a decade of bandwidth and wide scan angles past 60º from normal [10, 11, 12]. However, they are quite thick and have high cross-polarization when scanning in the D-plane. Furthermore, they are often fabricated using electronic discharge machining [10] or hand soldering a printed-circuit-board (PCB) grid together [13], which are expensive and time-consuming processes.

These fabrication challenges motivate the development of low-cost additively manufactured (i.e. 3D printed) Vivaldi arrays. One fabrication process involves 3D printing plastic and then electroplating the entire surface [14, 15]. Our group also found some success using this fabrication process in the past [16]. However, the electroplating process can be unreliable, especially in some of the critical areas such as near the feed. Furthermore, metal plated plastics typically have poor durability and temperature handling. Printing the antenna directly from metal using direct metal laser sintering (DMLS) is an attractive alternative to metal plated plastics [17]. A multitude of microwave components were fabricated using DMLS techniques [18]. Waveguide fed slots are particularly common antenna array elements [19, 20, 21]. Here, we propose using metal 3D printing to fabricate UWB arrays. This fabrication process is especially useful for research and development since customized designs can be cheaply built to order with short lead times. However, 3D metal printers have special design rules that need to be satisfied which are typically more stringent than plastic printers. Satisfying these design rules often requires significant modifications of the geometry and then re-optimization [22, 23, 24, 25]. Therefore, it is important to modify the standard Vivaldi geometry to be amendable to metal 3D printing, and then evaluate how these changes affect performance.

The vast majority of UWB antenna arrays utilize a square lattice, which is a natural geometry for integrating a vertical and horizontally polarized radiating element within a unit cell. However, it is well known that a triangular lattice offers 15.5% larger unit cell area for grating lobe free operation [26], which corresponds to a 0.6 dB larger gain for the same number of elements. Furthermore, a triangular lattice is often easier to fit within an arbitrary aperture shape on planar and/or curved surfaces [27]. These advantages have motivated the development of many triangular lattice arrays, most of which are narrowband. There are a few references to arrays with

C. Pfeiffer and T. Steffen are with Defense Engineering Corporation, Beavercreek, OH 45434, and also with US Air Force Research Laboratory, Wright-Patterson Air Force Base, OH 45433 (e-mail: carlpfei@umich.edu, tommy.steffen@teamdec.com).
J. Massman is with US Air Force Research Laboratory, Wright-Patterson Air Force Base, OH 45433 (e-mail: jeffrey.massman.5@us.af.mil)

greater than an octave bandwidth though. It is relatively straightforward to distribute a single linearly-polarized Vivaldi array on a triangular lattice by simply offsetting every other column [28]. However, extending this concept to a dual-polarizations is not trivial. The first dual-polarized UWB array on a triangular lattice was recently reported in [3], and achieves a 3:1 bandwidth ratio based on the PUMA architecture. Aside from slightly increased insertion loss and increased orthogonal port coupling, the triangular lattice PUMA performs similar to the square lattice version. An UWB triangular lattice Vivaldi array fabricated using electron discharge machining was also recently reported during the production of this manuscript [29].

Here, we report Vivaldi arrays 3D printed onto square and triangular lattices. This is the first UWB array fabricated using all metal 3D printing. The "Sub-Miniature Push-on, Micro" (SMPM) connectors are printed onto the radiating elements, which simplifies assembly and reduces cost compared to utilizing discrete RF connectors [16]. It is shown how to modify the Vivaldi geometry so that the design is both modular and satisfies the DMLS fabrication design rules. First, infinite array simulations are reported. Then, the arrays are fabricated and measurements are compared to simulations. Overall, the arrays have similar performance as previously published Vivaldi arrays [10, 12], but with simpler fabrication.

## II. Design

Dual-polarized Vivaldi arrays are commonly distributed in an egg crate geometry which separates the feed points of the $x$ and $y$ polarizations. A top view of Vivaldi designs arranged on square and triangular lattice egg crate geometries are shown in Fig. 1. The onset of grating lobes occurs at 20 GHz when scanning to 90° from normal for both the square and triangular lattice geometries (i.e. $\lambda_H = 15$ mm). Dual-polarized Vivaldi arrays require $x$ and $y$ directed arms to be orthogonal to each other, symmetric, and connected to neighboring elements to create the continuous transverse current that is required for ultra-wide bandwidth. The square lattice array naturally satisfies these conditions because the antenna elements can simply be arranged along the unit cell lattice. However, it was not obvious to us how to satisfy these conditions on a triangular lattice at the beginning of this project. Triangular lattices to date typically employ narrowband radiators that are isolated from each other such that the lattice geometry has minimal impact on the antenna design [30, 31, 32]. However, UWB radiators require strong coupling between neighboring elements to realize bandwidth ratios exceeding 3:1. Therefore, the antenna element design is directly influenced by the lattice geometry. The solution we devised is to place a slight vertical jog in the current flow between unit cells, as shown in Fig. 1(b). It should be emphasized that Vivaldi radiators are an extremely popular UWB array element and have been developed for several decades. However, all published versions have a square lattice despite the aforementioned advantages of a triangular lattice. Thus, the introduction of this geometry is novel in our opinion.

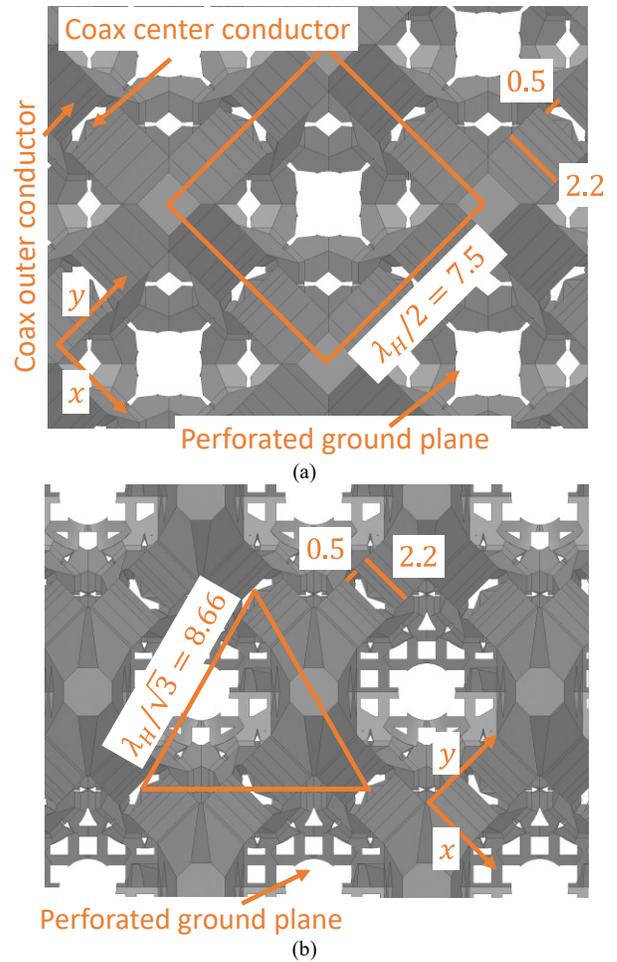

Fig. 1. Top view of the designed dual polarized Vivaldi arrays on square (a) and triangular (b) lattices. Listed dimensions are in mm.

Side views of the designed unit cells on square and triangular lattices are shown in Fig. 2. It is not possible to label all dimensions of the antennas here, but the most critical dimensions are shown. We printed the arrays from Titanium due to its 3D printing accuracy and decent conductivity ($\sigma = 1.82 \times 10^6$ S/m). The input SMPM connector is 3D printed onto the antenna such that the antenna can be measured directly after 3D printing since no discrete components need to be attached.

Several aspects of the element are implemented specifically for the direct metal laser sintering (DMLS) printing process. The printing process begins with a flat platform, and the part is built up in 30 μm thick layers. The manufacturer prints the antennas 'upside down' with the radiating tips attached to the build platform and the rest of the structure grows upwards from these tips, as shown in Fig. 3. The geometry is self-supporting in the sense that additional support structures between the build platform and the antenna are not necessary to hold up the antenna. Fabricating self-supporting geometry is more reliable since removing the unwanted support structures can be a manual and imprecise process. Since the part is self-supporting, everything must grow upwards and outwards. A rule of thumb for accurate fabrication is that the maximum angle from the vertical direction a part should grow at is roughly 45°. Thus, one design goal is to slowly sweep various geometries to minimize variation from one layer to the next.

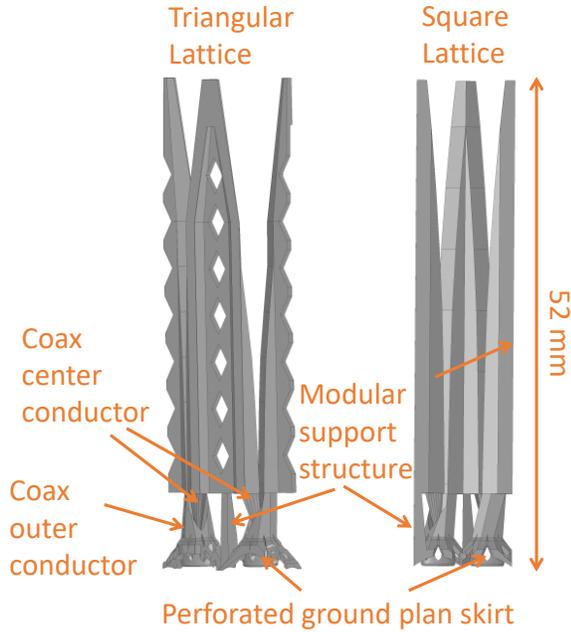

Fig. 2. Side view of the triangular and square lattice unit cells.

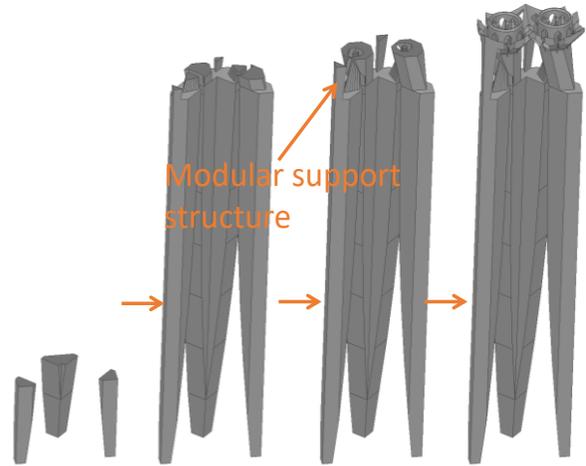

Fig. 3. Various stages of the fabrication process of a square lattice unit cell. The design is self-supporting such that the radiating tips are printed first and everything else grows upwards and outwards.

A novel feature of the antenna elements is the balun that converts the coaxial input connector into the balanced flared notch radiators. Conventional Vivaldi antennas are fed with a Marchand balun. However, Marchand baluns typically have a significant horizontal section that is not amendable to the flared angles required for self-supporting DMLS structures. Therefore, we chose to use a tapered transmission line balun such that the flared notch is excited by simply connecting the outer conductor and inner conductor of the coax feed to the two Vivaldi arms.

Another modification from classical Vivaldi antennas is the ground plane. The outer conductor of the coax feed is swept outwards at a near 45° angle to generate the ground plane. In contrast, conventional ground planes are horizontal which helps maximize the open volume of the Marchand balun and thus maximizes the bandwidth. Our ground plane skirt does slightly degrade the low frequency impedance match compared to an ideal horizontal ground plane. An advantage of our printed ground plane is the simple manufacturing since it is naturally electrically connected to the antenna elements. In contrast, it is common for traditional Vivaldi arrays to require hand soldering or conductive paste to connect the antenna elements to the ground plane.

The ground plane skirt is perforated with less than $\lambda_H/4$ diameter holes which helps reduce weight without sacrificing performance. In addition, these holes reduce material stress from large thermal gradients during the laser sintering process when the structure is printed, which in turn results in higher fabrication accuracy.

Since it is not possible to print very large arrays in a single run, it is important to ensure the design is modular so that subarrays can be combined to scale the array to arbitrary sizes. We designed an additional support structure into the antenna to improve its modularity for 3D printing. This modular support structure provides another connection between the Vivaldi arms and the ground plane skirt such that all features are mechanically connected. The structure allows for truncating the array along sections of the unit cell with low current density to minimize the impact of imperfect 'seams' between adjacent subarrays. For example, removing the modular support structure would disconnect the left-most arm in Fig. 3 from the rest of the structure such that the arm would be 'free-floating'. However, the support structure does degrade the low frequency performance. For example, the maximum VSWR without and with this structure is 2.5:1 and 2.9:1, respectively, around 3 GHz for broadside scan on the square lattice array.

III. INFINITE ARRAY SIMULATIONS

Different performance metrics for the two infinite array lattices are shown in Fig. 4 for broadside scan, as well as 60° from normal in the E, H, and D planes. Dashed vertical lines at 3 and 20 GHz denote the minimum and maximum operating frequencies. The VSWR is shown in Fig. 4(a) and Fig. 4(b). At broadside, the VSWR is below 2:1 for most frequencies from 4 to 20 GHz. The VSWR increases to 3:1 at lower frequencies around 3 GHz. The VSWR is below 3:1 for most frequencies across the band for scan angles out to 60° from normal. Overall, this performance is slightly worse than some of the previous state-of-the-art Vivaldi arrays arranged on square lattices that were fabricated using the more accurate and expensive electron discharge machining process [10, 12]. Further optimization of the unit cell could likely improve the low frequency performance.

Orthogonal port coupling refers to the power dissipated in the x-polarized antenna ports when all the y-polarized elements are excited, and vice versa. The orthogonal port coupling for the different scan directions and array lattices are shown in Fig. 4(c) and Fig. 4(d). The coupling is below -20 dB at broadside and below -10 dB at 60° scan angles. The coupling is most significant for scanning along the diagonal plane. Again, both the square and triangular lattices have coupling levels similar to previously published all metal Vivaldi arrays. It is worth noting that the orthogonal port coupling on the triangular lattice Vivaldi array is significantly lower than the triangular lattice PUMA array in [3], which is the only other dual-polarized UWB array fabricated on a triangular lattice to date. The difference in orthogonal port coupling between triangular lattice Vivaldi and PUMA arrays is likely due to the fact that Vivaldi arrays generally benefit from lower element-to-element coupling compared to low profile tightly coupled arrays.

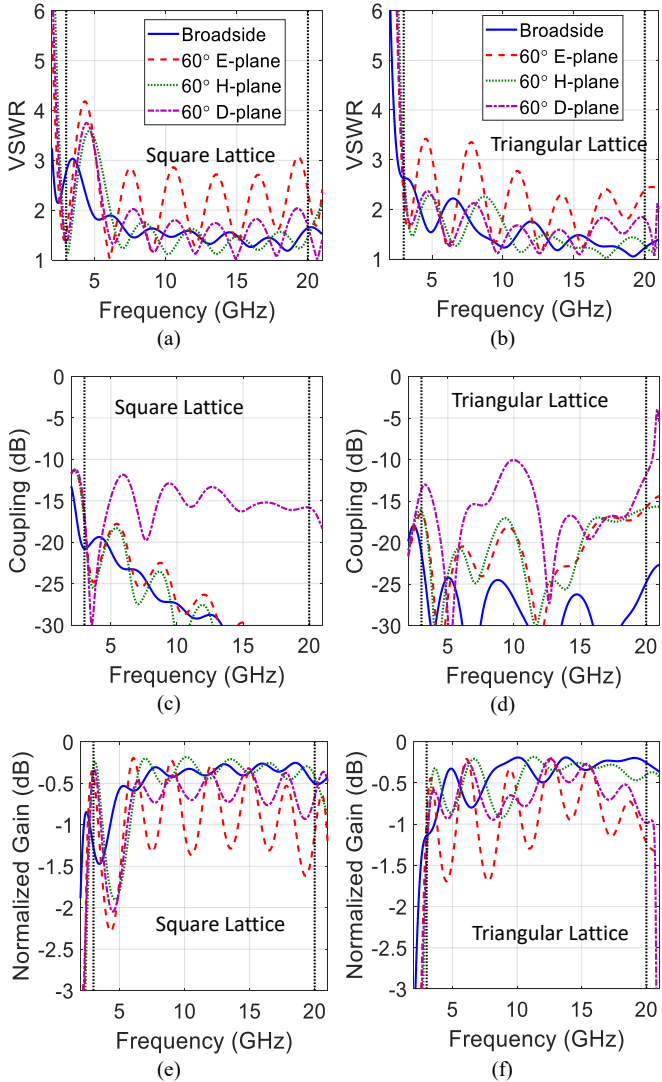

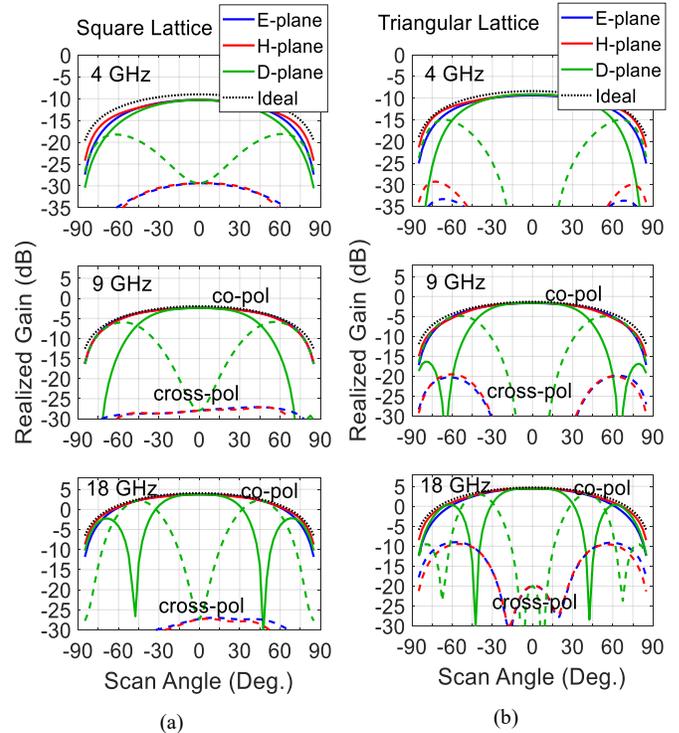

Fig. 5. Simulated embedded element patterns of an infinite array at 4, 9, and 18 GHz for the square (a) and triangular (b) lattices.

Fig. 4. Infinite array simulations of the square (a), (c), (e) and triangular (b), (d), (f) lattices when the array scans towards broadside and 60° in the E-, H-, and D-planes. (a) and (b) VSWR. (c) and (d) Orthogonal port coupling. (e) and (f) Normalized gain (ratio of power radiated to power incident).

Fig. 4(e) and Fig. 4(f) plot the normalized gain, which represents the fraction of incident power that is radiated by the array. In other words, the normalized gain is the product of the mismatch loss and radiation efficiency. Again, the gain shows the array performs quite well from 3-20 GHz. The gain is greater than -1 dB across most of the band and at most scan angles out to 60° from normal. The gain does dip down to -2 dB at some of the lower frequencies at wide scan angles. The simulated radiation efficiency is greater than 95% across the band (2-21 GHz) even though the metal conductivity is 30× lower than that of copper. This high radiation efficiency is due to the fact that the Vivaldi is not resonant, has relatively low peak current density, and only has a moderate electrical length of $3\lambda_H$ at the maximum operating frequency.

The embedded element patterns of the infinite array at 4, 9, and 18 GHz are shown in Fig. 5. The realized gain is plotted which corresponds to the gain multiplied by the mismatch loss. The dotted black line corresponds to the ideal embedded element with a $\cos(\theta)$ pattern and peak gain equal to $4\pi A/\lambda^2$, where $A$ is the unit cell area. The square lattice has low cross-polarization at all frequencies when scanning in the principal planes (E- and H-planes). However, the triangular lattice has moderate (-10 dB) cross-polarization levels at higher frequencies and wide scan angles in the principal planes. This higher cross-polarization is likely due to the sections of the radiating elements directed in the $\hat{x} - \hat{y}$ directions that connect neighboring antennas. The cross-polarization is high when scanning in the D-plane for both the square and triangular lattice arrays, which is typical for Vivaldi elements.

Overall, the Vivaldi arrays on square and triangular lattices perform similarly. The triangular lattice benefits from a 0.6 dB higher gain and can be more naturally truncated into a wide range of aperture shapes such as a rectangle, hexagon, or triangle. The square lattice has improved cross-polarization when scanning in the E- and H-planes. Square lattice Vivaldi antennas are generally easier to fabricate than triangular lattice versions when employing conventional techniques such as CNC machining, electron discharge machining, or PCB stacking. However, there is not a significant difference in ease of fabrication between the two lattice geometries when 3D printing.

## IV. Fabrication

The designed antennas are intended to be used in large arrays with 100's to 1000's of elements. However, smaller arrays are fabricated and their performance is compared to simulation to prove the concept. The fabricated square and triangular lattice arrays are shown in Fig. 6. The square lattice subarray has a square aperture with 24 dual-polarized elements, while the triangular lattice has a hexagonal aperture with 19 elements. The triangular lattice could easily have been truncated with a rectangular aperture, but a hexagon was chosen to highlight the aperture shape flexibility. Both arrays are 3D printed with titanium ($Ti_6Al_4V$) using the GE Additive Concept Laser M2, which can print parts up to 245mm x 245mm x 330mm in size. Many factors affect cost such as size, weight, and structural support removal time. To give a couple reference points, square and triangular lattice arrays from Fig. 6 weigh 97 g and 58 g, respectively. The overall costs of the square and triangular lattice arrays are $1540 and $1120 (USD), respectively. This translates into a price/element of $64 and $59 (USD), respectively. This low cost illustrates the utility of integrating the SMPM connector directly into the 3D printed antenna since commercial SMPM connectors on their own can cost roughly $60/element for dual-polarized designs. Furthermore, the cost of the antennas can be significantly reduced in the future by further reducing the weight/element, as well as increasing the array size to more efficiently utilize space on the build platform.

The 3D printed male SMPM connectors shown in Fig. 7 at the bottom of the antennas have the most critical dimensions. These connectors need to be precisely fabricated so that commercial female SMPM connectors can mechanically snap into the socket while also ensuring there is good electrical contact. A detent in the connector helps ensure a good connection is maintained if there is some vibration or stress on the input cables.

There are generally slight differences between the CAD models sent to the printer and the fabricated parts. Therefore, we printed several iterations of these connectors to compensate for these differences. For example, in the first iteration we simply printed the 3D CAD model of the ideal connector. However, the opening of the outer conductor was too small such that a commercial female connector could not fit. Therefore, we increased the size of this opening in the next round by 0.2 mm. After each iteration we plugged commercial connectors into the 3D printed parts and measured the reflection coefficient to evaluate how well the 3D printed connector performed. We also 'jiggled' the connectors to qualitatively evaluate how robust the connection is to misalignment error. Optical and 3D x-ray microscope images such as the one shown in Fig. 7(b) were also particularly helpful for diagnosing how accurately the connector was fabricated. Unfortunately, the precise transformation from CAD model to 3D printed part is generally printer dependent. Thus, it may be necessary to repeat this iterative process if a different printer is used in the future.

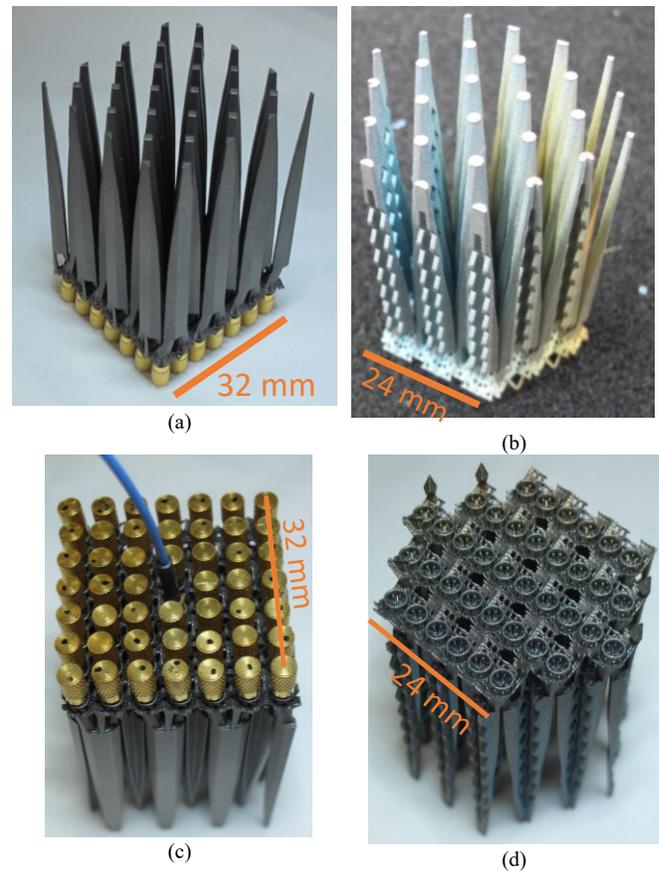

Fig. 6. Fabricated Vivaldi arrays. (a) Square lattice array with SMPM 50 Ω terminations connected to the ports. (b) Triangular lattice array. (c) Bottom view of square lattice array with coax cable feeding centermost element. (d) Bottom view of triangular lattice array.

This iterative process resulted in connectors with fabricated dimensions that are accurate enough for a good but not perfect connection to commercial SMPM connectors. Ideal SMPM connectors have a center conductor diameter of 0.3 mm. However, the measured center conductor diameter of the 3D printed part is 0.4 mm, which roughly corresponds to the minimum feature size of the standard resolution titanium printer we used. As shown in Fig. 7(b), the center conductor of the commercial connector flexes to allow the thicker-than-ideal 3D printed pins to fit inside. The center conductor of the 3D printed pin engages roughly 0.5 mm inside the center conductor of the commercial SMPM connector. In contrast, connections between two commercial SMPM connectors have an engagement around 0.8 mm. We found that printing connectors with larger than 0.5 mm engagement tended to damage the commercial female connector because the center conductor flexed too much to make room for the thick 3D printed pin. The reduced engagement in our design generally reduces the robustness to misalignment errors, but is still satisfactory for our purposes. For example, Fig. 7(c) plots measured the reflection coefficients of the 25 ports in the square lattice that are not located along the outermost edge. All unfed ports are terminated with 50 Ω loads. The overlapping curves in this figure illustrate the repeatability of these coaxial connections across the array. Note that this figure does not plot the active impedance match, and thus does not have a low reflection coefficient across the operating band.

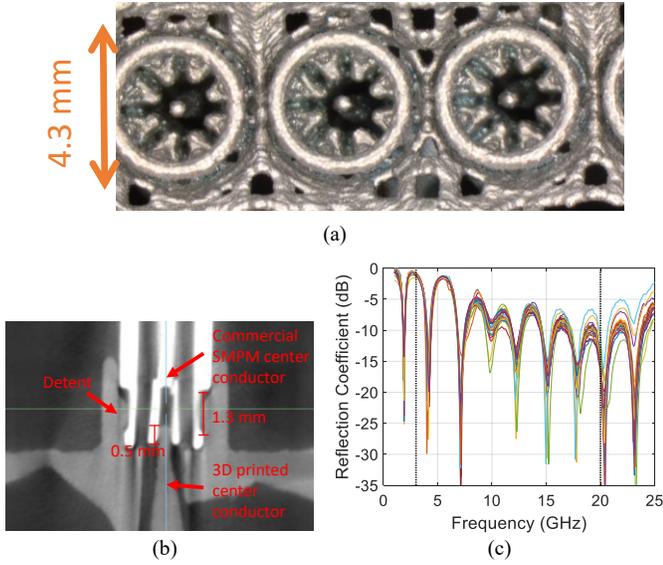

Fig. 7. Details of the printed SMPM connectors. (a) Zoomed in view of SMPM male connectors printed on the triangular lattice array. (b) Sideview x-ray image of a commercial SMPM female connector plugged into the 3D printed male connector. The commercial connector appears brighter in the image. (c) Measured reflection coefficients of the 25 ports in the square lattice that are not located along the outermost edge.

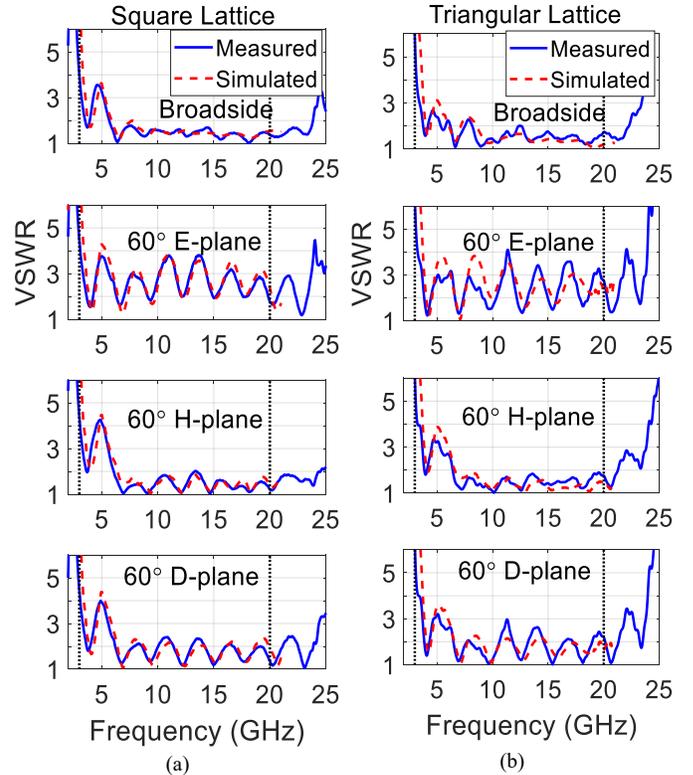

Fig. 8. Comparison of measured and simulated VSWR of the centermost element on the square (a) and triangular (b) finite arrays at various scan angles. The arrays are excited with a Gaussian amplitude taper with $w_0 = 1.1\lambda_H$ to minimize edge effects.

Higher resolution printers with minimum feature sizes around 0.15 mm are also available, which corresponds to a roughly 2x better resolution than the printer we used. However, higher resolution parts are generally more expensive and the maximum part size is smaller. In the future, these higher resolution printers could be used to improve the reliability of the SMPM connectors. Furthermore, they would allow the designs to be scaled up in frequency. It is likely our design could be scaled to operate up to 40 GHz using one of these printers with 2x better resolution. It should also be emphasized that significant investment is being put into 3D printing technologies which will likely improve the cost, maximum part sizes, and printing resolution of future designs. That said, a connectorized array would have other issues at frequencies as high as 40 GHz since we are not aware of any commercial RF connectors that are small enough to fit within a 40 GHz $\lambda/2$ lattice. A 40 GHz dual polarized square lattice would require two connectors to fit within 3.75 mm x 3.75 mm unit cell area.

## V. Measurements

We measured the active VSWR at the centermost element by measuring the reflection coefficient at that element as well as the transmission coefficient to all other elements in the array. All unused ports are terminated with 50 Ω loads. Then, the S-parameters are post processed to excite all elements with a given complex weight, and the power absorbed by the centermost element is noted. The array is excited with a Gaussian amplitude taper of $\exp(-r^2/w_0^2)$ across the aperture, where $r$ is the radial distance from the center of the array and $w_0$=17 mm ($1.1\lambda_H$) is the beam waist radius. This excitation more closely emulates a very large array with hundreds of elements since it reduces edge effects. The array is scanned to various angles by adding a linear phase shift across the elements. A comparison of the measured and simulated VSWR for the triangular and square lattice arrays are shown in Fig. 8. As expected, the finite array has a slightly larger VSWR than the infinite array simulations, especially at the lower operating frequencies. Nevertheless, measurements and simulations agree quite nicely which suggests the array is accurately fabricated.

Note that simulations also account for the same Gaussian amplitude taper. The simulated finite array performance is calculated by performing parametric sweeps of the infinite array active reflection coefficient at various scan angles. The S-parameter matrix is then extracted by taking an inverse fast Fourier transform (IFFT) of the infinite array active S-parameters. Additional details of the simulation process are provided in the Appendix.

A comparison between measured and simulated orthogonal port coupling is shown in Fig. 9. The same Gaussian amplitude taper with $w_0$=17 mm is applied to both measurements and simulations. Again, there is decent agreement between measurement and simulation. Coupling is strongest when scanning in the D-plane for both the triangular and square lattice arrays.

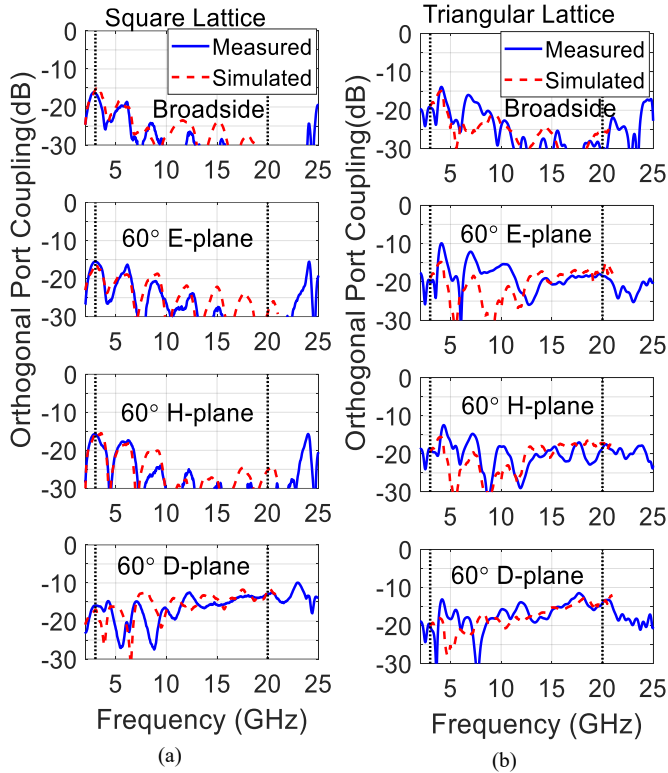

Fig. 9. Comparison of measured and simulated orthogonal port coupling of the centermost element on the square (a) and triangular (b) finite arrays at various scan angles. The arrays are excited with a Gaussian amplitude taper with $w_0 = 1.1\lambda_H$ to minimize edge effects.

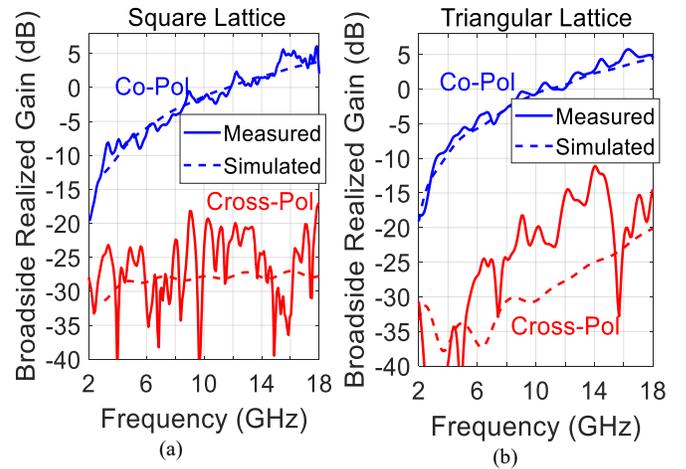

Fig. 10. Measured realized gain of the centermost element in the broadside direction on the square (a) and triangular (b) lattice arrays.

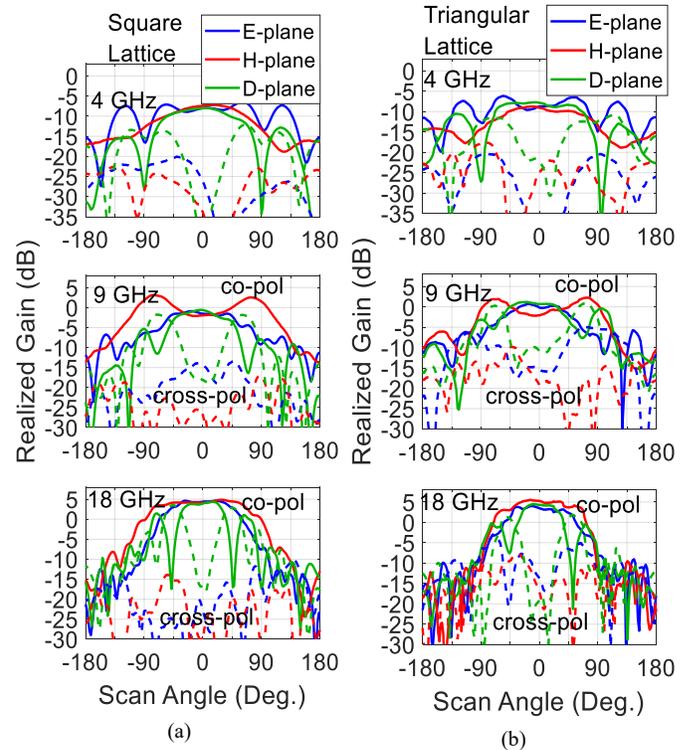

Fig. 11. Measured embedded element patterns of the centermost element on the square (a) and triangular (b) lattice arrays.

The realized gain in the broadside direction of the centermost element of the array is shown in Fig. 10. There is decent agreement between measurement and simulation. We employed the gain comparison method to characterize the fabricated antenna relative to a calibrated wideband horn antenna with known gain. The maximum measured frequency is limited to 18 GHz because this corresponds to the maximum operating frequency of the reference horn antenna. There is roughly 1.5 dB of ripple in the measured gain which is likely due to a combination of edge diffraction from the small arrays, reflections from walls in the measurement room, error in the calibrated gain table of the reference horn antenna, and imperfect 50 Ω termination loads of the unused array ports. Unfortunately, the measurement error in this setup is greater than the theoretical 0.6 dB gain difference between the square and triangular lattice arrays. Therefore, it is not possible to experimentally measure the gain advantage of the triangular lattice array over the square lattice array using our measurement system. The vertical and horizontal polarizations have similar gain values because the elements are perfectly symmetric.

The measured cross-polarization levels in the broadside direction are roughly 5 to 10 dB higher than in simulation. We suspect the high measured cross-polarization to be due to a loose connection in some of the SMPM 50 Ω terminations, which are shown in Fig. 6(c). For example, we ran a finite array simulation of the square lattice array, excited the centermost element, and adjusted the terminating impedance of one of the neighboring orthogonal polarized ports. The simulated cross polarized gain in the broadside direction at 9 GHz is -27 dB when the orthogonal polarized port is terminated in the ideal 50 Ω load. However, the cross-polarized gain increases to -14 dB when the neighboring port is open circuited. Thus, an improper termination of the unused ports can create a significant impact on the measured cross polarization. In contrast, the simulated co-polarization is not noticeably affected (~0.2 dB difference) by open circuiting the unused port.

The measured embedded element patterns at 4, 9, and 18 GHz are shown in Fig. 11. Similar to simulations in Fig. 5, the square lattice array maintains a low cross-polarization at all angles in the E- and H- scan planes. In contrast, the cross-polarization levels in the principal planes increase at wide scan angles and high frequencies for the triangular lattice array. The diagonal plane features high cross-polarization at all frequencies when scanning out to wide angles.

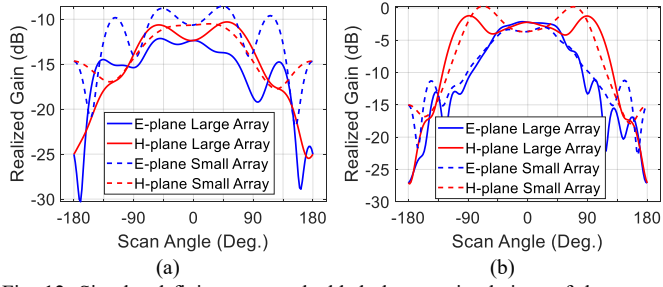

Fig. 12. Simulated finite array embedded element simulations of the square lattice array at (a) 4 GHz and (b) 9 GHz. The 'Small' array data corresponds to the 24 element array that was fabricated. The 'Large' array data corresponds to a 72 element array.

It should be noted that there are significant ripples in the patterns at the lower frequencies (4 GHz and 9 GHz). These ripples are due to diffraction from the edges of this small finite array. For example, Fig. 12 plots the simulated embedded element pattern of the centermost element in the square lattice array at 4 GHz and 9 GHz. Two different array sizes are simulated. The 'Small' array corresponds to the 24 element array that was fabricated, whereas the 'Large' array has 72 elements. The ripples in the 'Small' array agree well with the measured radiation patterns, and the ripples generally have a smaller amplitude than the 'Large' array. Of course, as the array size grows even more, the patterns will approach the infinite array case shown in Fig. 5.

## VI. Conclusion

We report the first all-metal additively manufactured Vivaldi arrays on square and triangular lattices. These particular arrays are designed to operate at 3-20 GHz and scan angles out to 60° from normal. It is shown how to modify the Vivaldi geometry so that the design is both modular and satisfies the DMLS fabrication design rules. An important feature of these arrays is the SMPM connector is directly printed with the antenna. The cost of these arrays with integrated connectors is roughly equal to the cost of commercial SMPM connectors alone. Furthermore, removing the additional step of soldering connectors at every element reduces cost and potentially improves reliability. Overall, the performance of the square and triangular lattice versions is similar, with the main difference being the triangular lattice has a max gain that is 0.6 dB higher than the square lattice for a given number of elements. However, the triangular lattice array does have higher cross-polarization levels when scanning in the principal planes. There is good agreement between measurement and simulation which illustrates the accuracy of the fabrication process. Additively manufactured arrays are particularly useful for research and development where the antenna can be customized for a given application, and then cheaply and rapidly manufactured.

Scaling these UWB arrays to higher operating frequencies and bandwidths is possible since higher resolution 3D printing services are currently commercially available. Scaling to lower frequencies is even more straightforward.

## VII. Appendix

Accurately simulating large finite arrays requires large computational resources. Therefore, many different methods have been developed that employ approximations to simplify the computational problem. Here, the finite array S-parameters are calculated by simulating an infinite array unit cell and sweeping the phase delay across the periodic boundary conditions on the sides of the simulation domain. The S-parameter matrix is found by taking the inverse Fourier transform of the active reflection coefficient vs scan angle. This process has been reported in general terms numerous times [33, 34]. However, we were unable to find explicit S-parameter matrix expressions for the cases of non-square lattices. The lack of these explicit expressions has led to some examples of misapplication of the method by some researchers. Therefore, we will review the method for calculating the S-parameter matrix from infinite array simulations.

Consider the coordinate systems shown in Fig. 1. The location of a given element in the array can be written as $\bar{R}_{n_1 n_2} = n_1 \bar{a}_1 + n_2 \bar{a}_2$, where $\bar{a}_1$ and $\bar{a}_2$ are the lattice vectors. The square lattice array is defined as,

$$[\bar{a}_1 \quad \bar{a}_2]_{square} = \frac{\lambda_H}{2} \begin{bmatrix} 1 & 0 \\ 0 & 1 \end{bmatrix} \quad (1)$$

and the triangular lattice is defined as,

$$[\bar{a}_1 \quad \bar{a}_2]_{traingle} = \frac{\lambda_H}{\sqrt{3}} \begin{bmatrix} -\sin\left(\frac{\pi}{12}\right) & 1/\sqrt{2} \\ \cos\left(\frac{\pi}{12}\right) & 1/\sqrt{2} \end{bmatrix} \quad (2)$$

For an array with $N$ elements and $N$ is assumed to be even, the active reflection coefficient can be written as,

$$\Gamma_{act}(\bar{k}_t) = \sum_{n_1=-\frac{N}{2}}^{\frac{N}{2}-1} \sum_{n_2=-\frac{N}{2}}^{\frac{N}{2}-1} S_{n_1 n_2, 0} \, e^{-j\bar{k}_t \cdot \bar{R}_{n_1 n_2}} \quad (3)$$

where $S_{n_1 n_2, 0}$ corresponds to the scattering parameter between the element located at $\bar{R}_{n_1 n_2}$ and the element at the center. The transverse wavenumber in the scan direction is given by $\bar{k}_t = k_x \hat{x} + k_y \hat{y}$. We decompose the transverse wavenumber into basis vectors, $\bar{k}_1$ and $\bar{k}_2$, such that $\bar{k}_t = m_1 \bar{k}_1 + m_2 \bar{k}_2$. Stipulating that $\bar{k}_t$ satisfies the following condition,

$$\begin{bmatrix} \bar{k}_1 \\ \bar{k}_2 \end{bmatrix} [\bar{a}_1 \quad \bar{a}_2] = \frac{2\pi}{N} \begin{bmatrix} 1 & 0 \\ 0 & 1 \end{bmatrix} \quad (4)$$

ensures that $\Gamma_{act}(\bar{k}_t)$ in (3) is a discrete Fourier transform of the scattering parameters. Note that $N\bar{k}_1$ and $N\bar{k}_1$ correspond to the standard reciprocal lattice primitive vectors. The scattering parameters can then be calculated by simply taking the inverse fast Fourier transform (IFFT) of (3),

$$S_{n_1 n_2, 0} = \frac{1}{N^2} \sum_{m_1=-\frac{N}{2}}^{\frac{N}{2}-1} \sum_{m_2=-\frac{N}{2}}^{\frac{N}{2}-1} \Gamma_{act}(\bar{k}_t) \, e^{j\bar{k}_t \cdot \bar{R}_{n_1 n_2}} \quad (5)$$

which can be written as,
$S_{n_1 n_2, 0} =$

$$\frac{1}{N^2} \sum_{m_1=-\frac{N}{2}}^{\frac{N}{2}-1} \sum_{m_2=-\frac{N}{2}}^{\frac{N}{2}-1} \Gamma_{act}(m_1 \overline{k}_1 + m_2 \overline{k}_2) \, e^{j2\pi/N \, (n_1 m_1 + n_2 m_2)} \quad (6)$$

where

$$\begin{bmatrix} \overline{k}_1 \\ \overline{k}_2 \end{bmatrix} = \frac{2\pi}{N} [\overline{a}_1 \quad \overline{a}_2]^{-1} \quad (7)$$

Inserting the lattice vectors for our square and triangular lattice from (1) and (2) into (7), gives the reciprocal lattice primitive vectors,

$$\begin{bmatrix} \overline{k}_1 \\ \overline{k}_2 \end{bmatrix}_{square} = \frac{4\pi}{N\lambda_H} \begin{bmatrix} 1 & 0 \\ 0 & 1 \end{bmatrix} \quad (8)$$

$$\begin{bmatrix} \overline{k}_1 \\ \overline{k}_2 \end{bmatrix}_{triangle} = \frac{4\pi}{N\lambda_H} \begin{bmatrix} -1/\sqrt{2} & 1/\sqrt{2} \\ \cos\left(\frac{\pi}{12}\right) & \sin\left(\frac{\pi}{12}\right) \end{bmatrix} \quad (9)$$

In the limit $N \gg 1$, the active reflection coefficient ($\Gamma_{act}(m_1 \overline{k}_1 + m_2 \overline{k}_2)$) can be easily calculated through computationally inexpensive unit cell simulations and sweeping the phase delay across the periodic boundary conditions.

To summarize, the scattering matrix for a triangular lattice array, for example, is calculated by simulating the unit cell at scan angles $m_1 \overline{k}_1 + m_2 \overline{k}_2$, where $m_{1,2} = -N/2$ to $N/2 - 1$ and $\overline{k}_{1,2}$ is given by (9). These active reflection coefficients ($\Gamma_{act}(m_1 \overline{k}_1 + m_2 \overline{k}_2)$) are inserted into (6) to calculate the scattering matrix ($S_{n_1 n_2,0}$). The accuracy of this process increases as the number of scan directions ($N$) is increased. A value of $N = 24$ is chosen here. Once the scattering matrix is found, it is straightforward to calculate metrics such as active VSWR and orthogonal port coupling for finite arrays excited with an arbitrary phase and amplitude distributions.

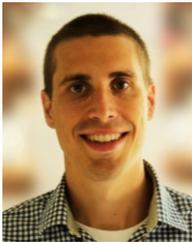
**Carl Pfeiffer** received the B.S.E., M.S.E. and Ph.D. degrees in electrical engineering from The University of Michigan at Ann Arbor, Ann Arbor, MI, USA, in 2009, 2011, and 2015 respectively.

He was a post-doctoral research fellow at The University of Michigan from April 2015 to March 2016. Since 2016 he has been with Defense Engineering Corp. as an onsite contractor for the Air Force Research Laboratory at Wright-Patterson Air Force Base, OH, USA. His research interests include phase arrays, engineered electromagnetic structures (metamaterials, metasurfaces, frequency selective surfaces), antennas, microwave circuits, plasmonics, optics, and analytical electromagnetics/optics.

Dr. Pfeiffer has served as a technical program committee member of EUCAP 2017 and has received top reviewer awards for IEEE Trans. Antennas Propag. in 2019 and 2020. He received the Kittyhawk AOC Research and Technology Development Award in 2020.

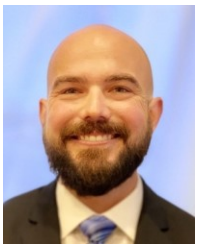
**Jeff Massman** received the B.S.E. and M.S.E. degrees in electrical engineering from United States Air Force Academy and Air Force Institute of Technology in 2008 and 2010, respectively.

He is currently the lead for the antenna and electromagnetic structures lab team with the Multiband Multifunction RF Sensing Branch of the Sensors Directorate Air Force Research Laboratory at Wright-Patterson Air Force Base, OH, USA and a PhD student at the Air Force Institute of Technology. His research interests include additive manufactured antennas, phased arrays, conformal frequency selective surfaces, and simultaneous transmit and receive devices.

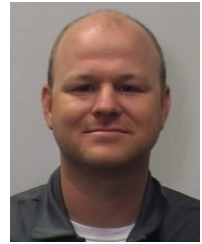
**Tommy Steffen** received his bachelors of science in Electrical Engineering from Wright State University at Dayton, Ohio in 2005.

He has worked as an RF Engineer for over 15 years of in the defense industry working with antennas and RF sensor systems. He began his career designing optically transparent antennas and small form factor, broadband apertures for ISR applications. He then transitioned to a Systems Engineer for electronic warfare sensors with a focus on high power active electronically scanned arrays using time delay and digital beamforming techniques. Since 2014 he has been with Defense Engineering Corporation, in Beavercreek, Ohio, collaborating with the Air Force Research Laboratory researching additively manufactured antennas and low-cost beamforming technologies. His research interests include low-cost radar sensors, digital beamforming, and additively manufacturing of broadband antenna arrays using plated plastics and metal 3D printing technologies.

Mr. Steffen has been a member of the Association of Old Crows since 2008 and was recipient of the 2015 AOC International Radio Frequency Award for outstanding achievement in development or application of RF systems, techniques, countermeasures, or counter-countermeasures.